\documentclass[%
 aip,
 amsmath,
 amssymb,
 reprint,%
]{revtex4-1}

\usepackage{amsfonts}
\usepackage{graphicx}
\usepackage[english]{babel}
\usepackage{color}
\usepackage{bm}
\DeclareMathAlphabet{\mathcal}{OMS}{cmsy}{m}{n}
\usepackage[version=3]{mhchem} 

\newcommand{\ket}[1]{\bigl | #1 \bigr \rangle}
\newcommand{\bra}[1]{\bigl\langle #1 \bigr|}
\newcommand{\braket}[2]{\langle #1 | #2 \rangle}

\definecolor{darkgreen}{RGB}{40,110,5}

\usepackage{ulem}  
\definecolor{orange}{rgb}{1,0.5,0}

\usepackage{dcolumn}

\usepackage[utf8]{inputenc}
\usepackage[T1]{fontenc}
\usepackage{mathptmx}
\usepackage{etoolbox}

\makeatletter
\def\@email#1#2{%
 \endgroup
 \patchcmd{\titleblock@produce}
  {\frontmatter@RRAPformat}
  {\frontmatter@RRAPformat{\produce@RRAP{*#1\href{mailto:#2}{#2}}}\frontmatter@RRAPformat}
  {}{}
}%
\makeatother

\begin{document}

\title{Perturbative calculation of non-linear response functions for multi-dimensional electronic spectra using non-Markovian quantum state diffusion. }

\author{Lipeng Chen}
\affiliation{Max Planck Institute for the Physics of Complex Systems, N\"{o}thnitzer Str 38, Dresden, Germany}
\email{lchen@pks.mpg.de}
\author{Doran I. G. Bennett}
\affiliation{Department of Chemistry, Southern Methodist University, PO Box 750314, Dallas, TX, USA}
\email{doranb@smu.edu}
\author{Alexander Eisfeld}
\affiliation{Max Planck Institute for the Physics of Complex Systems, N\"{o}thnitzer Str 38, Dresden, Germany}
\email{eisfeld@pks.mpg.de}

\begin{abstract}
We present a methodology for simulating multi-dimensional electronic spectra of molecular aggregates with coupling of electronic excitation to a structured environment using the stochastic non-Markovian quantum state diffusion (NMQSD) method in combination with perturbation theory for the response functions.
A crucial aspect of our approach is that we propagate the NMQSD equation in a doubled system Hilbert space, but with the same noise. 
We demonstrate that our approach  shows fast convergence  with respect to the number of stochastic trajectories, providing a promising technique for numerical calculation of two-dimensional spectra of large molecular aggregates.   
\end{abstract}
\maketitle

\section{Introduction}


Modern time-resolved nonlinear optical spectroscopies have expanded our understanding of the photophysics of molecular assemblies.\cite{ChemRevMukamel,MukamelTextBook,OKuhnTextBook,ValkunasBook,LPReview}
Two-dimensional (2D) electronic spectroscopy, the material response after interacting with three femtosecond laser pulses, is a particularly powerful probe of molecular excitons: 
2D spectra provide information about exciton-exciton interactions, dephasing, and relaxation processes.\cite{2DReview1,2DReview2,2DPaper1,2DPaper2,2DPaper3,2DPaper4,AkihitoCP}
Nevertheless, spectral congestion - even at low-temperatures - makes theoretical simulations indispensible for deciphering the dynamics encoded in the 2D spectra.

The key quantity in the simulation of 2D spectra is the third-order optically-induced polarization, which is related to third-order nonlinear response functions. \cite{MukamelTextBook} 
A common theoretical framework for simulating the third-order polarization of molecular aggregates is based on open-quantum system approaches, which propagate the reduced density matrix of the electronic system along different Liouville pathways to obtain nonlinear response functions. \cite{ChemRevMukamel,MukamelTextBook}
Of these approaches, those based on the Redfield or the modified Redfield equations are among the most popular:\cite{Redfield,MRedfield} their applicability, however, is restricted to the case of weak system-bath couplings and the Markovian approximation for the bath.
Alternatively, the hierarchy equation of motion (HEOM) \cite{HEOMReview1,HEOMReview2} and the quasiadiabatic path integral (QUAPI) \cite{QUAPINM1,QUAPINM2} provide numerically exact descriptions of the non-perturbative and non-Markovian dynamics.
However, while both methods have been widely used to simulate exciton dynamics and 2D spectra of molecular assemblies, \cite{HEOMApp1,HEOMApp2,HEOMApp3,QUAPIApp1,QUAPIApp2} they become numerically expensive for strong system-bath couplings, low temperatures, and large numbers of pigments.

An alternative to density matrix based methods is the {\it non-Markovian quantum state diffusion} (NMQSD) formalism, in which stochastic wavefunctions are propagated in the system Hilbert space and the density matrix is obtained from an average of these wavefunctions.\cite{NMQSD1,NMQSD2}
Over the years, several approaches have been developed to efficiently solve the NMQSD equation numerically,\cite{NMQSD3,Roden-ZOFE-PRL,HOPSPRL,HOPSDoran,Gao-MPS} so that simulations of excitation transport in large molecular aggregates containing thousands of pigments are now tractable. 
In these propagation schemes, importance sampling via the non-linear NMQSD equations are essential for efficient convergence with respect to the number of trajectories.
Within NMQSD, it is possible to obtain 2D spectra directly by including the femtosecond-pulses explicitly in the time-evolution and extracting the desired signal via phase-cycling, but the convergence with respect to the number of trajectories is slow compared to the calculation of expectation values.\cite{HOPS2DPC}

To overcome this problem, we develop in the present work an NMQSD equation in which the  response functions are obtained directly from a perturbative expansion with respect to interactions with the laser field. 
The crucial point of the scheme is to use the NMQSD formalism to propagate the combined  \textit{ket} and \textit{bra} states of the density matrix in a doubled electronic Hilbert space, but having the same noise.
Importance sampling  via the non-linear NMQSD equation introduces a coupling between the propagation of the \textit{bra} and \textit{ket} contributions.
We refer to NMQSD propagation in the doubled electronic Hilbert space as dyadic NMQSD, consistent with our previous treatment of linear absorption.\cite{ODHOPSLP}
Here, we solve the general dyadic NMQSD using a numerically efficient representation known as the Hierarchy of Pure States (HOPS).
Dyadic HOPS exhibits fast convergence with respect to the number of stochastic trajectories, and treats singly and doubly excited excitonic states in a unified manner, which is essential to account for ground state bleach (GSB), stimulated emission (SE), and excited state absorption (ESA) contributions to 2D spectra of molecular aggregates.  

This paper is organized as follows:
In section \ref{sec:Model}, we introduce the details of the molecular system, its interaction with electromagnetic pulses, and the general form of the response functions.
In section \ref{sec:ResponseInNMQSD}, we develop our method to calculate the response function using the NMQSD approach. 
We particularly emphasize the ability to use the non-linear NMQSD equation that ensures suitable convergence with respect to trajectories.
In section \ref{sec:example-calc}, we perform numerical calculation and demonstrate that with only 1000 trajectories spectra are already well-converged and discuss convergence trends in detail. Finally, we conclude in section \ref{sec:conclusions} with a summary and a brief outlook.
In Appendix~\ref{sec:LinearResponse}, we connect the first order (linear response)  of the present formalism to our previous calculations.\cite{ODHOPSLP}

\section{Molecular system, interaction with laser pulses and the quantities of interest \label{sec:Model} }

\subsection{Hamiltonian \label{sec:OpenQuantSysModel}}

We consider a molecular aggregate composed of $N$ interacting molecules, where each molecule is described by two electronic levels, the electronic ground state $|g_n\rangle$ and the electronic excited state $|e_n\rangle$, $n=1,\cdots,N$. 
The electronic Hamiltonian $\hat{H}_{\mathrm{ex}}$  can then be written as 
\begin{equation}
\hat{H}_\mathrm{ex}= \sum_n \epsilon_n \hat\sigma^\dagger_n \hat\sigma_n + \sum_n\sum_m V_{nm}\hat\sigma^\dagger_n\hat\sigma_m
\end{equation}
where  $\epsilon_n$ is the energy required to excite the $n$th molecule, $V_{nm}$ is the electronic coupling between excited molecules $n$ and $m$, and $\hat{\sigma}_n=\ket{g_n}\bra{e_n}$. 
In the case of 2D spectroscopy, we need  the common ground state $(|g\rangle=\prod_{n=1}^N|g_n\rangle)$, singly excited states $(|n\rangle=|e_n\rangle\prod_{m\neq{n}}|g_m\rangle=\hat{\sigma}_n^\dagger \ket{g})$, and doubly excited states  $(|nm\rangle=|e_n\rangle|e_m\rangle\prod_{k\neq{n,m}}|g_k\rangle=\hat{\sigma}_n^\dagger \hat{\sigma}_m^\dagger \ket{g}$, $n<m)$ of the molecular aggregate.

For each molecule there are additional interactions with internal and external nuclear degrees of freedom. In many cases of interest these interactions can be modelled by (infinite) sets of bosonic modes that couple linearly to the excitonic states. 
We denote these modes as \textit{environment} or \textit{bath}.
In this work we assume that each molecule has its own set of bath modes  so that the Hamiltonian of the bath can be written as 
\begin{equation}
\hat{H}_{\mathrm{B}}=\sum_{n=1}^N\sum_{\lambda}\hbar\omega_{n\lambda}\hat{b}_{n\lambda}^{\dagger}\hat{b}_{n\lambda}
\end{equation}
Here, $\hat{b}_{n\lambda}$ ($\hat{b}_{n\lambda}^{\dagger}$) is the annihilation (creation) operator of $\lambda$th bath mode of molecule $n$ with frequency $\omega_{n\lambda}$. 
The bath modes couple locally to their respective molecule.
The interaction Hamiltonian is then written as
\begin{equation}
\label{eq:H-int}
\hat{H}_{\mathrm{int}}=-\sum_{n=1}^N\hat{L}_n\sum_{\lambda}\kappa_{n\lambda}(\hat{b}_{n\lambda}^{\dagger}+\hat{b}_{n\lambda})
\end{equation}
where the coupling operator $\hat{L}_n$ acts in the system Hilbert space and is given by 
\begin{equation}
\hat{L}_n=\hat{\sigma}_n^\dagger \hat{\sigma}_n
\end{equation}
and $\kappa_{n\lambda}$ is the exciton-bath coupling strength of the mode $\lambda$ for molecule $n$, which is  specified by the bath spectral density of molecule $n$, $J_n(\omega)=\sum_{\lambda}|\kappa_{n\lambda}|^2\delta(\omega-\omega_{n\lambda})$. The latter is related to the bath-correlation function $\alpha_n(t)$ by
\begin{equation}
\alpha_n(t)=\int_0^{\infty}d\omega{J}_n(\omega)\left(\coth(\frac{\beta\omega}{2})\cos(\omega{t})-i\sin(\omega{t})\right)
\label{eq:alpha}
\end{equation}
with the inverse temperature $\beta=1/T$. 
We write the complete matter Hamiltonian as
\begin{equation}
    \label{eq:Hmatter}
    \hat{H}=\hat{H}_\mathrm{ex}+\hat{H}_\mathrm{B}+\hat{H}_\mathrm{int}
\end{equation}

\subsection{Interaction with Laser field}
\begin{figure}[tp]
    \includegraphics[width=0.45\textwidth]{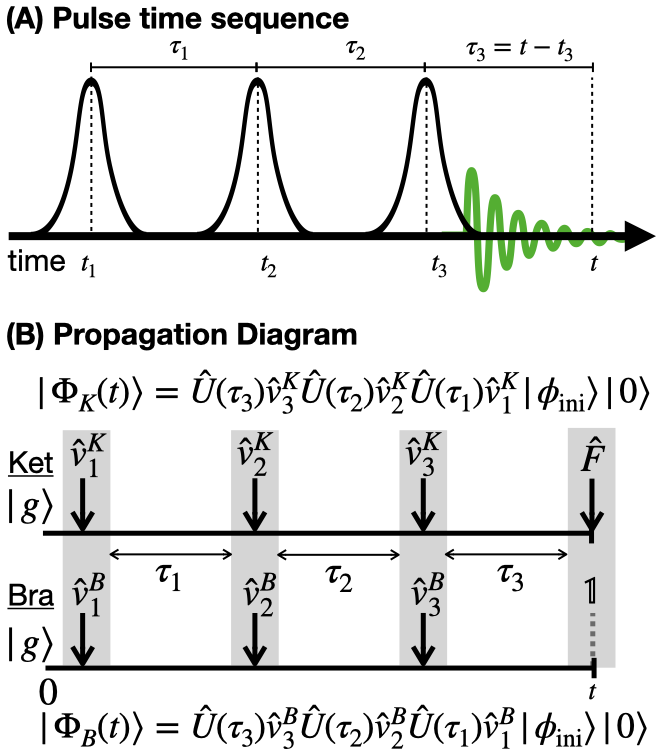}
    \caption{ (a) Sketch of the time sequence of the interaction with three short laser pulses, centered at times $t_1$, $t_2$, and $t_3$. 
    In the present work we focus on the impulsive limit, where the pulses are described by $\delta$-functions. After the third pulse a specific signal is detected at time $t_S$.
    The resulting expressions can be readily applied also to pulses with finite width.\cite{MukamelTextBook}
    Time intervals between two successive pulses are labeled by $\tau_j$ and indexed according to the index of the first of the two pulses. 
    (b) Symbolic representation of the calculation of the response function according to the formalism of section \ref{sec:ReponeDoubleSpace}. 
    }
    \label{fig:time_sequence}
\end{figure}

In a 2D spectroscopy experiment, the system interacts with three laser pulses at controlled inter-pulse delay times. 
The field-matter interaction Hamiltonian $\hat{H}_{\mathrm{F}}(t)$ is defined as \cite{MukamelTextBook} 
\begin{equation}
\hat{H}_F(t)=-\hat{\boldsymbol{\mu}}\cdot\mathbf{E}(\mathbf{r},t),
\end{equation}
where, $\hat{\boldsymbol{\mu}}$ is the total transition dipole operator 
\begin{equation}
\label{eq:hatmu}
    \hat{\boldsymbol{\mu}}=\sum_{n=1}^N\boldsymbol{\mu}_n (\hat\sigma_n +\hat\sigma_n^\dagger) 
\end{equation}
and $\boldsymbol{\mu}_n$ the transition dipole moment of molecule $n$.

 The electric field is given by 
\begin{equation}\label{EF}
\mathbf{E}(\mathbf{r},t)=\sum_{a=1}^3\mathbf{e}_aE_{a}(t-t_a)e^{i\mathbf{k}_a\cdot\mathbf{r}-i\omega_a(t-t_a)}+\mathrm{H.c.}
\end{equation}
with $\mathbf{e}_a$, $\mathbf{k}_a$, $\omega_{a}$, $E_a(t)$, and $t_a$ denoting the polarization unit vector, the wave vector, the carrier frequency, the envelope, and the central time of the $a$th pulse, respectively. 
In general there can also a different number of pulses.

We can write the total Hamiltonian $\hat{H}_{\mathrm{tot}}$ as 
\begin{eqnarray}
\begin{aligned}
\hat{H}_{\mathrm{tot}}=&\hat{H}+\hat{H}_{\mathrm{F}}(t) 
\end{aligned}
\end{eqnarray}
with $\hat{H}$ given in Eq.~(\ref{eq:Hmatter}).

\subsection{Response functions\label{sec:ResponseFunct}}

In this section we present a general notation for non-linear optical response functions that provide a clear connection to the NMQSD formalism. 
Here, we use a general notation that that is appropriate for arbitrary orders of perturbation theory. 
In section \ref{sec:example-calc}, we specify to 2D spectroscopy.

\subsubsection{Perturbation theory for the full density matrix}

In the following we denote the time-evolution operator of the system without the electromagnetic field as
\begin{equation}
    \hat{U}(\tau)=\hat{U}_\tau=e^{-i \hat{H} \tau}
\end{equation}
where $\hat{H}$ given in Eq.~(\ref{eq:Hmatter}).
We also introduce the abbreviation 
\begin{equation}
\hat{V}_j=\hat{H}_F(t_j)
\end{equation}
for the interaction with field at time $t_j$.

We are interested in correlation functions (that we loosely call response functions) which depend on the order of interactions with the electric field. 
We use a condensed notation to track the generic correlation functions of the form
\begin{equation}
\label{eq:R-general}
     R\!\!\left[{\footnotesize \begin{array}{c}
     \tau_1,\dots,\tau_M\\
      v^{K}_1,\dots,v^{K}_M      \\
       v^{B}_1, \dots, v^{B}_M  
     \end{array} } \right] =  \mathrm{Tr}\Big\{ \hat{F}  \,
     \hat{\rho}^{(M)}\!\!\!\left[{\footnotesize \begin{array}{c}
     \tau_1,\dots,\tau_M\\
      v^{K}_1,\dots,v^{K}_M      \\
       v^{B}_1, \dots, v^{B}_M  
     \end{array} } \right] \Big\}
\end{equation}
where the expectation value of an observable $\hat{F}$, in our case the polarization $\mathbf{\mu}$, is calculated with respect to a density matrix that is obtained in $M^\mathrm{th}$ order of perturbation theory. 
In Eq.~(\ref{eq:R-general}) the parameters in the rectangular brackets  track the order of interactions influencing the \textit{ket} (row 2) vs \textit{bra} (row 3) time evolution associated with a particular time-correlation function. 
The first row contains the intervals $\tau_i=t_{i+1}-t_i$ between interaction at time $t_i$ and the following interaction.
The last interval $\tau_M$ contains the final time $t$ of the evolution, i.e., $\tau_M=t-t_M$.
The sequence of operators in the second and third row identify the operators acting on the \textit{bra} and \textit{ket} side (respectively) at each interaction time and thereby define a specific response function.
The $M^\mathrm{th}$ order density matrix is recursively defined by
\begin{equation}
\label{eq:rho_recursiv}
    \hat{\rho}^{(j)}\!\!\!\left[{\footnotesize \begin{array}{c}
     \tau_1,\dots,\tau_j\\
      v^{K}_1,\dots,v^{K}_j      \\
       v^{B}_1, \dots, v^{B}_j  
     \end{array} } \right]=\hat{U}({\tau_j})\, \hat{v}^K_j\, \hat{\rho}^{(j-1)}\!\!\!\left[{\footnotesize \begin{array}{c}
     \tau_1,\dots,\tau_{j-1}\\
      v^{K}_1,\dots,v^{K}_{j-1}      \\
       v^{B}_1, \dots, v^{B}_{j-1}  
     \end{array} } \right] \big(\hat{v}^B_j\big)^\dagger \hat{U}^\dagger({\tau_j})
\end{equation}
and 
\begin{equation}
 \hat{\rho}^{(0)}=   \ket{\phi_\mathrm{ini}}\bra{\phi_\mathrm{ini}}\otimes \rho_\mathrm{env}. 
\end{equation}
In Eq.~(\ref{eq:rho_recursiv}) the operators  ${v}_j^\mathrm{K}$ act always on the \textit{ket} side and the operators ${v}_j^\mathrm{B}$ always on the \textit{bra} side of $\hat{\rho}$.
An important constraint is that for each pair ${v}_j^\mathrm{K}$,  ${v}_j^\mathrm{B}$, with the same index $j$ one of the corresponding operators is the unit operator (which we denote by $\openone$).
Note that for better readability we will sometimes omit the `operator hats' on the operators ${v}_j^{\mathrm{K/B}}$, as it has already been done above.
Explicitly we have
\begin{align}
\begin{array}{c}
    {v}_j^\mathrm{K}=\hat{V}_j\\
    {v}_j^\mathrm{B}=\openone
\end{array}
\quad \mathrm{or}
\quad
\begin{array}{c}
    {v}_j^\mathrm{K}=\openone\\
    {v}_j^\mathrm{B}=\hat{V}_j
\end{array}
\end{align}

In the following, we introduce a short-hand notation where we abridge 
\begin{equation}
    \hat{\rho}^{(M)}=\hat{\rho}^{(M)}\!\!\!\left[{\footnotesize \begin{array}{c}
     \tau_1,\dots,\tau_M\\
      v^{K}_1,\dots,v^{K}_M      \\
       v^{B}_1, \dots, v^{B}_M  
     \end{array} } \right] \
\end{equation}
by omitting all arguments when it is clear which correlation function is being considered or when we consider generic correlation functions.

\section{Calculation of the non-linear response function using NMQSD \label{sec:ResponseInNMQSD}}
\subsection{The NMQSD formalism} 

 For the open quantum system model $\hat{H}=\hat{H}_\mathrm{ex}+\hat{H}_\mathrm{B}+\hat{H}_\mathrm{int}$ as given in section \ref{sec:OpenQuantSysModel} and 
 for a factorized initial state ${\hat{\rho}}_{\mathrm{ini}}=|\phi_{\mathrm{ini}}\rangle\langle\phi_{\mathrm{ini}}|\otimes\hat{\rho}_{\mathrm{env}}$, the expectation value of any system operator $\hat{F}$ can be obtained as \cite{NMQSD1,NMQSD2} 
\begin{equation}
\label{eq:normalized_ExpectationVal}
\langle\hat{F}(t)\rangle \equiv \mathrm{Tr}[\hat{F}{\hat{\rho}}(t)]=\mathcal{M}\left[\frac{\langle\phi(t,\mathbf{z}^{*})|\hat{F}|\phi(t,\mathbf{z}^{*})\rangle}{\langle\phi(t,\mathbf{z}^{*})|\phi(t,\mathbf{z}^{*})\rangle}\right].    
\end{equation}
where $\mathcal{M}[\cdots]$ denotes ensemble average over stochastic wave function $|\phi(t,\mathbf{z}^{*})\rangle$ obtained by the normalizable (non-linear) NMQSD equation \cite{NMQSD1,NMQSD2}
\begin{equation}\label{eq:dotphit_NMQSD}
\begin{split}
\partial_t|\phi(t,\mathbf{z}^{*})\rangle=&-i\hat{H}_{\mathrm{ex}}|\phi(t,\mathbf{z}^{*})\rangle\\
&+\sum_n \hat{L}_n\, \zeta_{n}(t,\mathbf{z})|\phi(t,\mathbf{z}^{*})\rangle\\
&-\sum_n \left(\hat{L}_n^{\dagger}-\langle\hat{L}_n^{\dagger}\rangle_t\right)\int_0^t\mathrm{d}s\,\alpha_n(t-s)\frac{\delta{|}\phi(t,\mathbf{z}^{*})\rangle}{\delta{z}_{s,n}^{*}}
\end{split}
\end{equation} 
where $\mathbf{z}$ comprises a set of complex Gaussian stochastic processes $z_{t,n}^{*}$ with mean $\mathcal{M}[z_{t,n}]=0$, and correlations $\mathcal{M}[z_{t,n}z_{s,m}]=0$ and $\mathcal{M}[z_{t,n}z_{s,m}^{*}]=\alpha_n(t-s)\delta_{nm}$.
Here $\alpha_n(t)$ is the correlation function  of the environment, defined in Eq.~(\ref{eq:alpha}).
These processes enter via
$\zeta_{n}(t,\mathbf{z})=z_{t,n}^{*}+\int_0^tds \,\alpha_n^{*}(t-s)\langle\hat{L}_n^{\dagger}\rangle_s$, where the expectation values $\langle\cdot\rangle_t$ are calculated using the normalized state
$|\phi(t,\mathbf{z}^{*})\rangle/\sqrt{\langle\phi(t,\mathbf{z}^{*})|\phi(t,\mathbf{z}^{*})\rangle}$.

For completeness we mention that beside the non-linear NMQSD equation (\ref{eq:dotphit_NMQSD}), there exists also a {\it linear} NMQSD formulation where the non-linear terms $\langle \hat{L}_n^\dagger \rangle$ are dropped in Eq.~(\ref{eq:dotphit_NMQSD}) and in $\zeta_n(t,\mathbf{z})$ and expectation values are calculated as $\langle \hat{F} \rangle (t)=\mathcal{M}[\langle\phi(t,\mathbf{z}^{*})|\hat{F}|\phi(t,\mathbf{z}^{*})\rangle]$.
However, the linear NMQSD equation converges slowly with the number of trajectories except for the case of weak system environment coupling or very short propagation times.

\subsection{Reformulation of the response function equations\label{sec:ReponeDoubleSpace}} 

Our aim is now to formulate the response function in a way that can be used together with the above  non-linear NMQSD equation.
We first note that in the derivation of the NMQSD equation  the finite temperature initial state of the environment has been transformed to the vacuum states by shifting the contributions of temperature to the bath-correlation function. 
Therefore, the following derivation is done for the initial state $\hat{\rho}^{(0)}= \ket{\phi_\mathrm{ini}}\bra{\phi_\mathrm{ini}}\otimes \ket{\mathbf{0}}\bra{\mathbf{0}}$.

We introduce $\ket{\Phi_\mathrm{B}(t)}$ and $\ket{\Phi_\mathrm{K}(t)}$ which represent the evolution of the \textit{bra} and \textit{ket} contributions, respectively. 
We can then write the $M^{th}$ order density matrix as 
\begin{equation}
\hat{\rho}^{(M)}(t)=\ket{\Phi_\mathrm{K}(t)} \bra{\Phi_\mathrm{B}(t)}
\end{equation} 
where
\begin{align}
    \ket{\Phi_\mathrm{B}(t)}= &(e^{-i \hat{H} \tau_M} \hat{v}_{M}^{\mathrm{B}})\cdots  (e^{-i \hat{H} \tau_1}\hat{v}_{1}^{\mathrm{B}})\ket{\phi_\mathrm{ini}}\ket{\mathbf{0}}
    \\
     \ket{\Phi_\mathrm{K}(t)}= &(e^{-i \hat{H} \tau_M} \hat{v}_{M}^{\mathrm{K}})\cdots  (e^{-i \hat{H} \tau_1}\hat{v}_{1}^{\mathrm{K}})\ket{\phi_\mathrm{ini}}\ket{\mathbf{0}}
\end{align}
and   the last interval $\tau_M=t-t_M$ contains the time $t$. 
In this notation, the response function $R(t)$, an abbreviation for $R(t)=R^{(M)}(\tau_1,\dots,\tau_M)$, becomes
\begin{align}
    R(t)=&\mathrm{Tr}\{\hat{F} \ket{\Phi_\mathrm{K}(t)} \bra{\Phi_\mathrm{B}(t)} \}
    \\
    =&
   \bra{\Phi_\mathrm{B}(t)} \hat{F} \ket{\Phi_\mathrm{K}(t)}.
   \label{eq:BFK_full}
\end{align}
Both $\ket{\Phi_\mathrm{B}(t)}$ and $\ket{\Phi_\mathrm{K}(t)}$ can be expanded  with respect to coherent states of the bath
\begin{align}
\ket{\Phi_\mathrm{B/K}(t)}=
   \int dM(\mathbf{z}) \ket{\mathbf{z}}\braket{\mathbf{z}}{\Phi_\mathrm{B/K}(t)}= &\int dM(\mathbf{z}) \ket{\mathbf{z}}\ket{\phi_\mathrm{B/K}(t,\mathbf{z}^*)}
\end{align}
where $\ket{\phi_\mathrm{B/K}(t,\mathbf{z}^{*})}$ are states in the `system' Hilbert space only and $dM(\mathbf{z})= \Pi_{n\lambda} d^2 z_{n\lambda}\frac{e^{-|z_{n\lambda}|^2}}{\pi} $.
Inserting this expansion into Eq.~(\ref{eq:BFK_full})
\begin{align}
    R(t)=&
  \int dM(\mathbf{z})  \int dM(\mathbf{z}') \braket{\mathbf{z}}{\mathbf{z}'}  \bra{\phi_\mathrm{B}(t,\mathbf{z}^*)} \hat{F} \ket{\phi_\mathrm{K}(t,\mathbf{z}^{'*})}
\end{align}
and using the `reproducing property' of  coherent states, we  obtain the important result
\begin{align}
\label{eq:R_as_Int_dM(z)}
    R(t)=&
  \int dM(\mathbf{z})    \bra{\phi_\mathrm{B}(t,\mathbf{z}^*)} \hat{F} \ket{\phi_\mathrm{K}(t,\mathbf{z}^{*})}
\end{align}
 where the {\it bra} and the {\it ket} now evolve with the same coherent states $\mathbf{z}$.

Introducing a state
\begin{equation}
\label{eq:def_double}
\ket{\widetilde\psi(t,\mathbf{z}^*)}
= 
\begin{pmatrix}
\ket{\phi_\mathrm{B}(t,\mathbf{z}^{*})}\\
\ket{\phi_\mathrm{K}(t,\mathbf{z}^{*})}
\end{pmatrix}
\end{equation}
in a doubled {\it `system'} Hilbert space,
we can write
\begin{equation}
\label{eq:R_as_Int_dM(z)_double}
R(t)=
\int dM(\mathbf{z}) \bra{\widetilde\psi(t,\mathbf{z}^*)}
\widetilde{F}
\ket{\widetilde\psi(t,\mathbf{z}^*)}
\end{equation}
with
$\widetilde{F}=\begin{pmatrix}
0 & \hat{F}\\
0 &0
\end{pmatrix}
$.
These formulas are the starting point for the dyadic NMQSD approach,  similar to the doubling used for the case of a quantum state diffusion unravelling of Lindblad equations.\cite{QSDdoublespace}

\subsection{Dyadic NMQSD}
The dyadic NMQSD equations use the construction of the response functions in the doubled `system' Hilbert space to time-evolve the combined \textit{bra} and \textit{ket} states.
 We can introduce a state $\ket{\widetilde \Psi(t)}$ that lives in the Hilbert space $(\mathcal{H}_S \otimes  \mathcal{H}_S) \otimes \mathcal{H}_B$ (note that the `bath' Hilbert space is not doubled) and obeys
\begin{equation}
 \ket{\widetilde\psi(t,\mathbf{z}^*)}=\int dM(\mathbf{z})\ket{\mathbf{z}}\braket{\mathbf{z}}{\widetilde\Psi(t)}.
\end{equation}
where the left hand side is the state introduced in Eq.~(\ref{eq:def_double}).
For the corresponding time evolution we can write
\begin{equation}
\ket{\widetilde\Psi(t)}=\widetilde{\mathcal{U}}(\tau_M) \widetilde{V}_M \cdots \widetilde{\mathcal{U}}(\tau_1) \widetilde{V}_1 \ket{\widetilde\Psi(0)} 
\end{equation}
with an initial state
\begin{equation}
\label{eq:psi_ini_double}
\ket{\widetilde\Psi(0)}=
\begin{pmatrix}
\ket{\phi_\mathrm{ini}}\\
\ket{\phi_\mathrm{ini}}
\end{pmatrix}
\otimes
\ket{\mathbf{0}},
\end{equation}
a time-evolution operator
\begin{equation}
\widetilde{\mathcal{U}}(\tau)=e^{-i \widetilde{H} \tau},
\end{equation}
and a Hamiltonian
\begin{equation}
\widetilde{H}=\widetilde{H}_S
+ \hat{H}_B
+ \sum_n \widetilde{L}_n\sum_\lambda \kappa_{n\lambda} (\hat{b}_{n\lambda}^{\dagger}+ \hat{b}_{n\lambda})
\end{equation}
where the $\kappa_{n\lambda}$ are the same as in Eq.~(\ref{eq:H-int}),
\begin{equation}
\widetilde{H}_S=\begin{pmatrix}
\hat{H}_S & 0\\
0 & \hat{H}_S
\end{pmatrix},
\,\,
\widetilde{V}_j= \begin{pmatrix}
\hat{v}_j^\mathrm{B} & 0\\
0 & \hat{v}_j^\mathrm{K}
\end{pmatrix}, \,\, \mathrm{and} \,\,
\widetilde{L}_n=\begin{pmatrix}
\hat{L}_n & 0\\
0 & \hat{L}_n
\end{pmatrix}.
\end{equation}
The response function (Eq.~(\ref{eq:R_as_Int_dM(z)_double})) is then evaluated as an average over trajectories which obey the nonlinear NMQSD equation~(Eq. ~\eqref{eq:dotphit_NMQSD}) in the doubled system Hilbert space. 
There are many formally equivalent dyadic NMQSD propagation schemes; we will discuss one explicit propagation scheme in Section \ref{sec:numeric_prop}, below.

While the \textit{bra} and \textit{ket} states are not directly coupled within the dyadic non-linear NMQSD equations, they both contribute to the norm of the dyadic wave function and cannot be propagated independently.
For example, note that the expectation values $ \langle \tilde{L}^\dagger_n\rangle_t$, that appears in  Eq.~(\ref{eq:dotphit_NMQSD}), are calculated using the dyadic wave function in the doubled system Hilbert space, where 
 \begin{align}
   \langle \tilde{L}^\dagger_n\rangle_t =& \frac{\bra{\widetilde\psi(t,\mathbf{z}^*)}\widetilde{L}^\dagger_n\ket{\widetilde\psi(t,\mathbf{z}^*)}}{\braket{\widetilde\psi(t,\mathbf{z}^*)}{\widetilde\psi(t,\mathbf{z}^*)}}\\
   =&
   \frac{\bra{\phi_\mathrm{K}(t,\mathbf{z}^*)}\hat{L}^\dagger_n\ket{\phi_K(t,\mathbf{z}^*)}+\bra{\phi_\mathrm{B}(t,\mathbf{z}^*)}\hat{L}^\dagger_n\ket{\phi_B(t,\mathbf{z}^*)}}{||\phi_\mathrm{K}(t,\mathbf{z}^*)||^2+||\phi_\mathrm{B}(t,\mathbf{z}^*)||^2 }
 \end{align}
and 
\begin{equation}
    || \phi(t,\mathbf{z}^*)||^2= \braket{\phi_\mathrm{}(t,\mathbf{z}^*)}{\phi_\mathrm{}(t,\mathbf{z}^*)}.
\end{equation}

We note that for the linear dyadic NMQSD equation, the bra and ket states are not coupled, however, these equations show poor convergence in parameter regimes beyond weak system-bath coupling.

\subsection{Summary of the numerical propagation scheme \label{sec:numeric_prop}}

In the present work the response function is calculated in the following way:
We start with the system part of the initial state Eq.~(\ref{eq:psi_ini_double}), which lives in the doubled system Hilbert space and which reads $\ket{\widetilde{\phi}_\mathrm{ini}}=\big(\ket{\phi_\mathrm{ini}},\ket{\phi_\mathrm{ini}}\big)^T$.
This state is not normalized.
On this state we act with $\widetilde{V}_1$. 
We then propagate using the non-linear (but unnormalized) NMQSD  during the time interval $\tau_1$. 
Then we act with the second interaction $\widetilde{V}_2$ and continue the propagation during time-interval $\tau_2$.
We repeat this until the end of the last time interval and then calculate the expectation value of $\widetilde{F}$ for each individual trajectory.

According to Eq.~(\ref{eq:normalized_ExpectationVal}) we normalize each trajectory before taking the average.
Here some care is necessary. 
The normalization should take care of the change of norm caused by the un-normalized NMQSD propagation but it includes in addition the norm changes due to the interactions $\widetilde{V}_1, \dots ,\widetilde{V}_M$.
To keep these physically relevant changes of the norm we multiply by these norm changes.
In detail: Let us denote the state (in doubled system Hilbert space) before the $j$-th interaction by $\widetilde{\psi}(t_j,z^{*})$, where $t_j$ is the time of the $j$-th interaction.
We define
\begin{equation}
    \mathcal{I}_j = \frac{|| \widetilde{V}_j\widetilde{\psi} (t_j,z^{*})||^2}
    {||\widetilde\psi(t_{j+1},z^{*})||^2}
\end{equation}
We then have for the `response function'
\begin{equation}
    R^{(M)}(z)=\Big(\prod_{j=1}^M \mathcal{I}_j \Big) \bra{\widetilde{\psi}(t,z^{*})}\tilde{F}\ket{\widetilde{\psi}(t,z^{*})}, 
\end{equation}
and obtain the final result by averaging over trajectories
\begin{eqnarray}
R^{(M)}= \mathcal{M}_z \big\{ R^{(M)}(z) \big\}.
\end{eqnarray}
We summarize this procedure in Fig.~\ref{fig:time_sequence}(b). We could have also used the {\it normalized} non-linear NMQSD equation, with an appropriate change to the normalization factors at the end.

\subsection{HOPS for solving the NMQSD propagation }
The NMQSD equation Eq.~(\ref{eq:dotphit_NMQSD}) is not particularly suitable for a direct numerical implementation, because of the functional derivative with respect to the stochastic processes.
Numerically convenient schemes can be derived when the bath-correlation function $\alpha_n(t)$ is expanded as a finite sum of exponentials,
\begin{equation}
\label{eq:alpha-exps}
    \alpha_n(t)\approx\sum_{j=1}^{N_\mathrm{modes}} p_{nj}e^{-w_{nj}t}
\end{equation}     
    with $w_{nj}=\gamma_{nj}+i\Omega_{nj}$.
In many applications of practical interest, the required number of `modes' $(N_\mathrm{modes})$ is small.
For the interpretation of such modes see for examples. \cite{AlexPMode1,AlexPMode2} 
A powerful, but approximate, scheme that is based on Eq.~(\ref{eq:alpha-exps}) is the so called `zeroth order functional expansion' ZOFE.\cite{NMQSD3,Roden-ZOFE-PRL} 
In the present work we employ the numerically exact hierarchy of pure states (HOPS) \cite{HOPSPRL}

\begin{eqnarray} \label{NonLinHOPSEq}
\begin{aligned}
\partial_t|\psi^{(\mathbf{k})}(t,\mathbf{z}^{*})\rangle=&\Big(-i\hat{H}_{\mathrm{ex}}- \mathbf{k}\cdot\mathbf{w}+\sum_n \hat{L}_n\,\zeta_{n}(t,\mathbf{z})\Big)|\psi^{(\mathbf{k})}(t,\mathbf{z}^{*})\rangle\\
&+\sum_{n}\hat{L}_n \sum_j k_{nj}p_{nj}|\psi^{(\mathbf{k}-\mathbf{e}_{nj})}(t,\mathbf{z}^{*})\rangle  \\
&-\sum_{n}\left(\hat{L}_n^{\dagger}-\langle{\hat{L}}_n^{\dagger}\rangle_t\right)\sum_j |\psi^{(\mathbf{k}+\mathbf{e}_{nj})}(t,\mathbf{z}^{*})\rangle.
\end{aligned}
\end{eqnarray}
 $\mathbf{w}=\left\lbrace{w}_{1,1},\cdots,w_{N,J}\right\rbrace$, and $\mathbf{k}=\left\lbrace{k}_{1,1},\cdots,k_{N,J}\right\rbrace$ with non-negative integers $k_{nj}$. 
The vector $\mathbf{e}_{nj}=\left\lbrace{0},\cdots,1,\cdots,0\right\rbrace$ is one at the $(n,j)$th position and is zero otherwise. 
The relevant contribution to perform calculations of expectation values is the zeroth order element, i.e.
\begin{equation}
    \ket{\phi(t,\mathbf{z}^*)}=|\psi^{(\mathbf{0})}(t,\mathbf{z}^{*})\rangle.
\end{equation}

The HOPS consists of an infinite set of coupled equations, which must be truncated at a finite hierarchy for numerical calculations. 
In this work, we use a simple triangular truncation condition for the hierarchy: $|\mathbf{k}|\leq\mathcal{K}$. 
More advanced truncation schemes are discussed in Ref.~ \onlinecite{TrunHOPS}.
It is also possible to use an adaptive algorithm to reduce the size of the hierarchy \cite{HOPSDoran} or use a matrix product state representation. \cite{Gao-MPS}

\section{Example calculations}\label{sec:example-calc}
One reason for developing the present perturbative approach is the large number of trajectories required to converge the non-perturbative approach of Ref.~\onlinecite{HOPS2DPC}.
Therefore, in the following exemplary calculations we focus in particular on the convergence with respect to the number of trajectories.

\subsection{Model system}
Here, we perform  calculations for a dimer ($N=2$) consisting of identical monomers (i.e.~$\epsilon_n=\epsilon$) with  parallel  transition dipoles  ($\boldsymbol{\mu}_n=\boldsymbol{\mu}$).
For the bath-correlation functions we choose a single exponential
\begin{equation}
\label{eq:single_exp}
 \alpha_n(t)=\alpha(t)=pe^{-i\Omega{t}-\gamma t}, \quad \quad t\ge 0  
\end{equation}
with $\gamma=0.25$ and the vibrational  frequency $\Omega$ as the unit of energy. 
Then Eq.~(\ref{eq:single_exp}) can be interpreted as a weakly damped vibrational mode at zero temperature,\cite{AlexPMode1} which requires a non-Markovian treatment. 
Below, we consider two values for the coupling strength that lead to qualitative different 2D spectra: the intermediate coupling case  and the strong coupling case ($p=0.5$ and $1.8$, respectively in units of $\Omega^2$).
For the interaction between the monomers we use $V=0.3$.

\subsection{The various 2D spectra}
In the following, we consider the third order response functions that contribute to 2D electronic spectroscopy.
For the three time intervals we adopt the commonly used notation
\begin{equation}
    \tau_1\equiv\tau, \quad \quad \tau_2\equiv T, \quad \quad \tau_3 \equiv t.
\end{equation}
We present plots for the ground state bleaching (GSB), stimulated emission (SE) and excited state absorption (ESA) signals. 
\begin{equation}
\begin{split}
S_{\mathrm{GSB}}(\omega_{\tau},T,\omega_t)=&\phantom{-((}S_3^{(-)}(\omega_{\tau},T,\omega_t)+S_4^{(+)}(\omega_{\tau},T,\omega_t)\\
S_{\mathrm{SE}}(\omega_{\tau},T,\omega_t)=&\phantom{-((}S_2^{(-)}(\omega_{\tau},T,\omega_t)+S_1^{(+)}(\omega_{\tau},T,\omega_t) \\
S_{\mathrm{ESA}}(\omega_{\tau},T,\omega_t)=&-\big(S_5^{(-)}(\omega_{\tau},T,\omega_t) +S_6^{(+)}(\omega_{\tau},T,\omega_t)\big)
\end{split}
\end{equation}
with
\begin{equation}
\label{eq:S+-}
S^{(\pm)}_\ell(\omega_{\tau},T,\omega_t)=\mathrm{Re}\int_0^{\infty}\int_0^{\infty}\mathrm{d}t\mathrm{d}\tau \ \ r_\ell(\tau,T,t)e^{\pm i\omega_{\tau}\tau} e^{i \omega_tt}
\end{equation}
These expressions emerge after applying the rotating wave approximation and phase matching conditions.\cite{MukamelTextBook,ChemRevMukamel}
The functions $r_\ell$ appearing in Eq.~(\ref{eq:S+-}) are specific response functions that are evaluated for $F=\mathbf e \cdot \boldsymbol{\hat{\mu}_{-}}$, and contain non-hermitean interaction operators  $\hat{V}_j^{\pm} = - \hat{\boldsymbol{\mu}}_{\pm} \mathbf{E}_j^{\pm}$, where
\begin{eqnarray}
\label{eq:mu+-}
    \hat{\boldsymbol{\mu}}_{+}=\sum_n \vec{\mu}_n \hat\sigma_n^\dagger, &\quad\quad\quad& \hat{\boldsymbol{\mu}}_{-}=\sum_n\vec{\mu}_n \hat\sigma_n, \\
    \mathbf{E}_j^{+}=\mathbf{e}_j, &\quad\quad\quad & 
    \mathbf{E}_j^{-}=\mathbf{e}_j^{*}
    \label{eq:E+-}
\end{eqnarray}
In Table~\ref{tab:r1-r6} we summarize the operators that enter into the calculation of the response functions $r_1$ to $r_6$. 
We take all fields to be identical and polarized parallel to the transition dipole moments of the molecules.

\begin{table}[pht]
\begin{tabular}{>{\centering}m{0.7cm}|| >{\centering}m{1cm} >{\centering}m{1cm}  |  >{\centering}m{1cm}  >{\centering}m{1cm} |>{\centering}m{1cm} >{\centering\arraybackslash}m{1cm}}
     &  $v_1^\mathrm{K}$ & $v_1^\mathrm{B}$  & $v_2^\mathrm{K}$ & $v_2^\mathrm{B}$ & $v_3^\mathrm{K}$  & $v_3^\mathrm{B}$ \\ [0.1cm]
     \hline
     \\ [-0.2cm]
   $r_1$  & $\hat{V}_1^+$ & $\openone$ & $\openone$ & $\hat{V}_2^{+}$ & $\openone$ &$\hat{V}_3^{-}$ 
   \\ [0.1cm]
    $r_2$  &   $\openone$ &$\hat{V}_1^+$ &  $\hat{V}_2^{+}$ & $\openone$ & $\openone$ &$\hat{V}_3^{-}$ 
  \\ [0.1cm]
    $r_3$  &   $\openone$ &$\hat{V}_1^+$ &  $\openone$ & $\hat{V}_2^{-}$ & $\hat{V}_3^{+}$ &$\openone$
  \\ [0.1cm]
    $r_4$  &   $\hat{V}_1^{+}$ &$\openone$ &  $\hat{V}_2^{-}$ & $\openone$ & $\hat{V}_3^{+}$ &$\openone$
  \\ [0.1cm]
    $r_5$  &   $\openone$ &$\hat{V}_1^{+}$ &  $\hat{V}_2^{+}$ & $\openone$ & $\hat{V}_3^{+}$ &$\openone$
  \\ [0.1cm]
    $r_6$  &   $\hat{V}_1^{+}$ &$\openone$ &  $\openone$ & $\hat{V}_2^{+}$ & $\hat{V}_3^{+}$ &$\openone$
\end{tabular}
\caption{\label{tab:r1-r6}The operators used in the calculation of the response functions $r_1$, \dots $r_6$. 
Here $\hat{V}_j^{\pm} = - \hat{\boldsymbol{\mu}}_{\pm} \mathbf{E}_j^{\pm}$ with $\hat{\boldsymbol{\mu}}_{\pm}$ and $\mathbf{E}_j^{\pm}$ given in Eqs.~(\ref{eq:mu+-}) and (\ref{eq:E+-}).}
\end{table}

\begin{figure*}
\begin{center}
  \includegraphics[width=17.5cm]{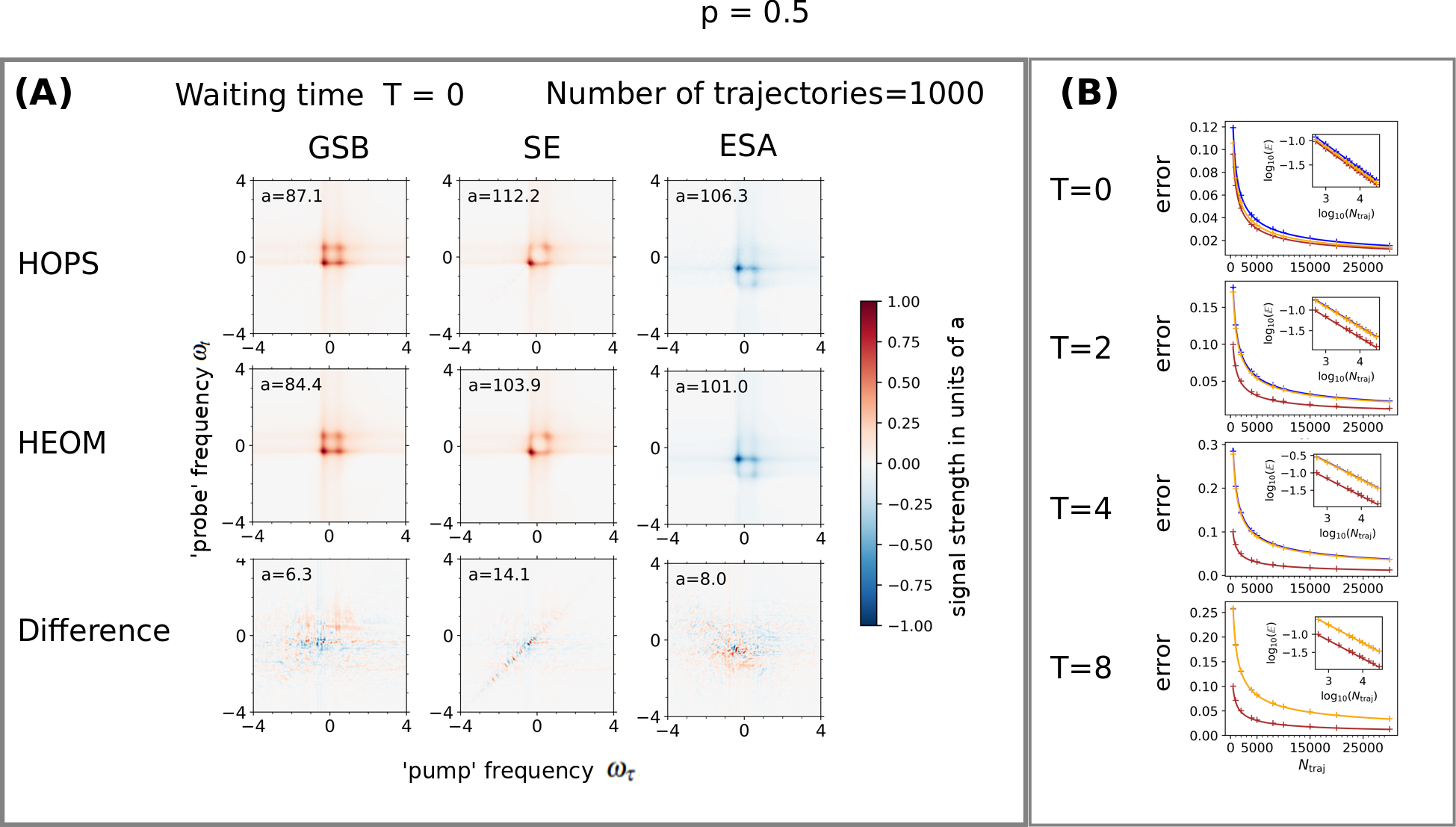}
   \end{center}
    \caption{Two-dimensional spectra and their convergence properties for a dimer with two identical monomers and intermediate system-bath coupling ($p=0.5$).
  The other parameters are $\gamma=0.25$ and $V=0.3$, where the energies are in units of $\Omega$.
   For $p=0.5$ the truncation conditions for HOPS  are $\mathcal{K}=10$ for the GSB and SE, and $\mathcal{K}=11$ for the ESA.
  (A) GSB, SE and ESA spectra at waiting time $T=0$. 
  Upper row: HOPS calculations with $N_\mathrm{traj}=1000$ trajectories.
  Second row: HEOM calculations as reference.
  Third row: point-wise difference $S_\mathrm{HOPS}(\omega_\tau,\omega_t) -S_\mathrm{HEOM}(\omega_\tau,\omega_t)$ between the HOPS and the HEOM spectra. 
  Note that we have normalized all spectra to the maximal value $a=\max [S(\omega_\tau,\omega_t)]$, which is indicated in each panel.
   (B) Convergence with the number of trajectories for different waiting times $T$.
   In each panel the error measure, defined in appendix \ref{sec:error}, is shown as function of number of trajectories.
   The crosses are the numerical data obtained from a bootstrapping procedure (described in appendix \ref{sec:error}) and the solid lines are curves that follow a $1/\sqrt{N_\mathrm{traj}}$ scaling. GSB: brown, SE: orange, ESA: blue.
    \label{fig:Weak}
    }
\end{figure*}

\begin{figure*}
\begin{center}
   \includegraphics[width=17.5cm]{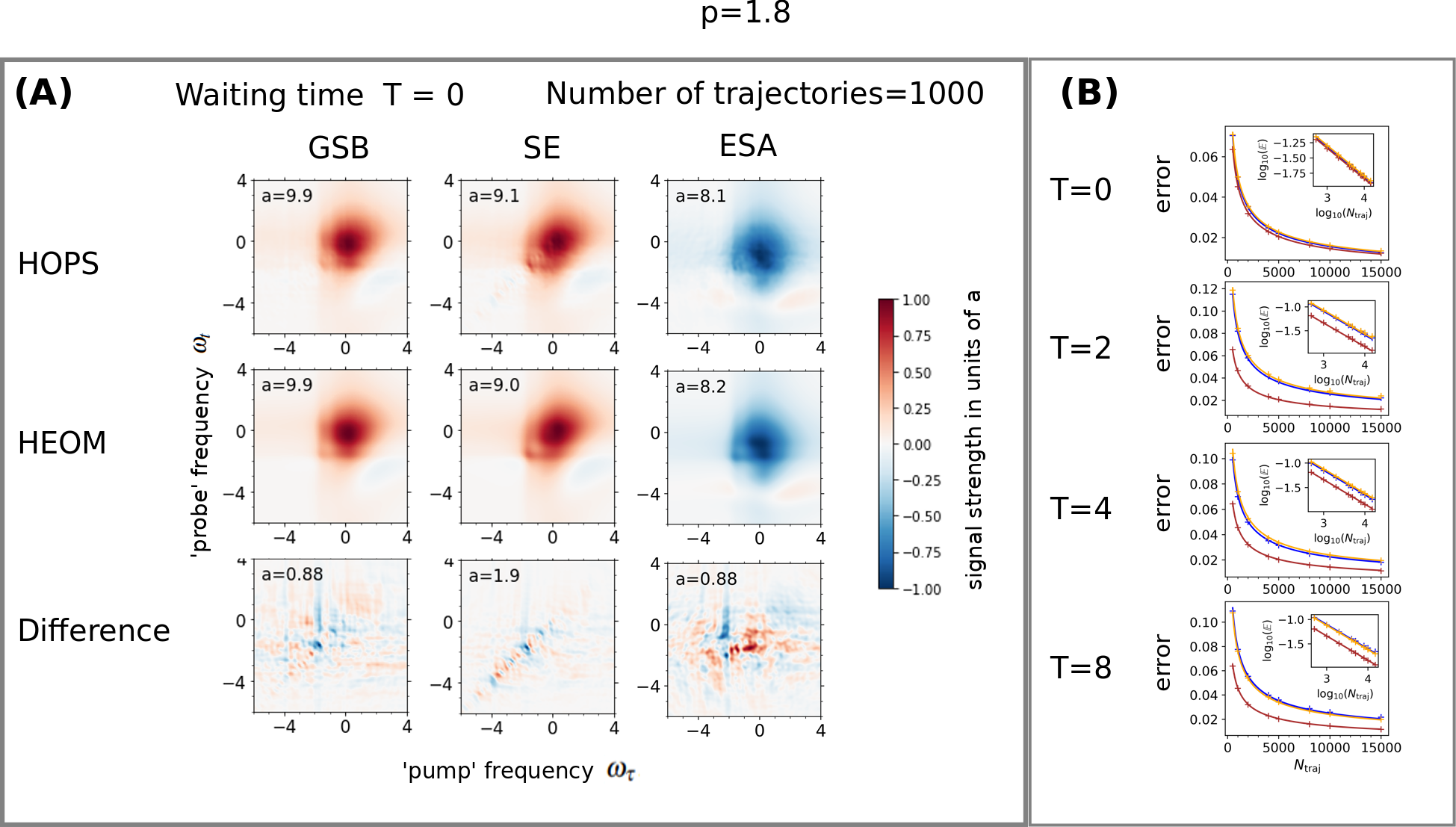}
   \end{center}
    \caption{Same as Fig.~\ref{fig:Weak} but for the strong system-bath coupling regime ($p=1.8$). The truncation conditions for HOPS  are $\mathcal{K}=15$ for the GSB and SE, and $\mathcal{K}=20$ for the ESA.
    }
    \label{fig:Strong}
\end{figure*}

\subsection{Calculations}

In Fig.~\ref{fig:Weak} we present the intermediate coupling case, $p=0.5$. 
Fig.~\ref{fig:Weak}~(A)  shows in the upper row the GSB, SE, and ESA spectra obtained for T=0 using $N_\mathrm{traj}=1000$ trajectories.
We compare our HOPS spectra to reference calculations, performed with the HEOM method\cite{HEOMReview1,HEOMReview2}  (Fig.~\ref{fig:Weak}A, middle row). 
We see that the HOPS spectra reproduces the relevant features from HEOM.
To see in  detail the deviations of the HOPS spectra from the HEOM ones, we show the point wise difference between the  HOPS spectrum and the reference HEOM spectrum (Fig.~\ref{fig:Weak}A, bottom  row).
From this one sees that the maximal differences are around 10\% of the peak signal.
For GSB and ESA the fluctuations are spread around a large region in the vicinity of the signal.
For the SE spectrum the fluctuations are largest along the diagonal.

To investigate the convergence with the number of trajectories we introduce a measure $\mathbb{E}$ for the integrated difference (details are given in appendix \ref{sec:error}). 
This measure of the difference  for the GSB, ES, and ESA (Fig.~\ref{fig:Weak}A) are given by $\mathbb{E}=0.069$, $\mathbb{E}=0.076$ and $\mathbb{E}=0.085$, respectively. 
From this we see that $\mathbb{E}\lesssim 0.08$ corresponds to quite good agreement.
In Fig.~\ref{fig:Weak}B, we plot a non-parametric error estimate (calculated via bootstrapping, Appendix \ref{sec:error}) with respect to the number of trajectories.
The SE (orange) and ESA (blue) errors are  very similar and both are larger than the GSB error (brown).
At waiting time $T=0$ all three errors are comparable; upon increasing the waiting time the error of the GSB signal remains essentially unchanged, while the error of the SE and ESA signal increases.
For the shown waiting times $T\approx 4$ the SE and ESA error curves also remain largely unchanged and even decrease slightly. 
For each waiting time, the error follows the expected $1/\sqrt{N_\mathrm{traj}}$ scaling (shown as a solid line), which can be clearly observed in the inset showing a double logarithmic scale.

Fig.~\ref{fig:Strong} shows analogous results for the strong coupling case.
In particular, there is again the $1/\sqrt{N_\mathrm{traj}}$ scaling of the error and the error does not increase for waiting times $T\geq 2$.

\section{Conclusions}\label{sec:conclusions}
In the present work, we have developed a framework to simulate multidimensional electronic spectra of molecular aggregates  using the stochastic non-linear formalism to directly calculate perturbative response functions of arbitrary order. 
Our approach propagates a dyadic wave function which combines \textit{ket} and \textit{bra} states in a doubled system Hilbert space subjected to a common noise.

Numerical simulations for a dimer system coupled to a structured environment demonstrate that our theory  has favorable convergence properties with respect to the number of stochastic trajectories. 
It should be noted that the new formalism developed here needs many fewer stochastic trajectories to obtain converged 2D spectra when compared to those calculated by a non-perturbative  phase-cycling scheme.\cite{HOPS2DPC} 
As compared to density matrix based methods,  the non-linear NMQSD method  propagates vectors instead of matrices,  individual simulations with different noise trajectories are trivially parallel and, furthermore, it is consistent with adaptive basis \cite{HOPSDoran} and tensor contraction \cite{Gao-MPS} approaches recently developed for HOPS.
Our theory thus offers a promising technique to simulate 2D spectra of large molecular aggregates. 
This is especially the case for the description of the excited state absorption contribution to 2D spectra where a large number of doubly excited excitonic states are involved. 
Furthermore, it is straightforward to account for the effect of the static disorder induced by the inhomogeneity of the solvent environment by simply sampling excitonic parameters from a certain distribution for each stochastic trajectory.
Our theory can be readily applied to simulate higher-order response functions, for example, fifth-order 3D signals, which are a powerful tool to reveal multi-step energy transfer processes. \cite{LHCII3D}

\begin{acknowledgements}
We thank Jacob K.\ Lynd for proofreading.
LPC  acknowledges support from the Max-Planck Gesellschaft via the MPI-PKS visitors program. 
 AE acknowledges support from the DFG via a Heisenberg fellowship (Grant No EI 872/10-1).
 DIGB acknowledges support from Robert A. Welch Foundation (Grant N-2026-20200401) and the US National Science Foundation CAREER Award under Grant CHE-2145358. 
\end{acknowledgements}

\section*{Data availability}
Further data that support the findings of this study are available from the corresponding author upon reasonable request.

\appendix

\clearpage
\section{Linear response}\label{sec:LinearResponse}

It is instructive to also consider linear response within the present formalism  to elucidate  the normalization with respect to the state in doubled Hilbert space.
The linear response function as it appears in the calculation of absorption is defined as 
\begin{equation}
\label{eq:Resp_LinAbs}
    R^{(1)}(t)=\mathrm{Tr}\left\lbrace(\mathbf{e}^{*}\cdot\hat{\boldsymbol{\mu}}_{-})e^{-i\hat{H}t}(\mathbf{e}\cdot\hat{\boldsymbol{\mu}}_{+})|g\rangle\langle{g}|\hat{\rho}_{\mathrm{B}}e^{i\hat{H}t}\right\rbrace.
\end{equation}

\subsection{Expression using the formalism of section \ref{sec:ResponseInNMQSD}}
Within the dyadic NMQSD formalism we  write the linear response function as 
\begin{equation}
    R^{(1)}(t)= \mathcal{M} \Big[\mathcal{I}_1 \langle\tilde{\psi}(t,\mathbf{z}^{*}) \vert \tilde{F} \vert \tilde{\psi}(t,\mathbf{z}^{*}) \rangle \Big],
\end{equation}
where 
\begin{equation*}
\widetilde{F}=\begin{pmatrix}
0 & (\mathbf{e}^{*}\cdot\hat{\boldsymbol{\mu}}_{-}) \\
0 & 0 
\end{pmatrix},
\quad
 \mathcal{I}_1 = \frac{|| \widetilde{V}_1\widetilde{\psi} (t_1,\mathbf{z}^{*})||^2}
    {||\widetilde\psi(t,\mathbf{z}^{*})||^2},
\end{equation*}
\begin{equation*}
\widetilde{V}_1=\begin{pmatrix}
\openone & 0 \\
0& (\mathbf{e}\cdot\hat{\boldsymbol{\mu}}_{+}) 
\end{pmatrix},
\quad \mathrm{and} \quad 
\ket{\widetilde{\psi}(t_1,\mathbf{z}^{*})}=\begin{pmatrix}
\ket{g}\\
\ket{g}
\end{pmatrix}.
\end{equation*}
This corresponds to the first-order response function with $\hat{v}^\mathrm{K}_1=\mathbf{e}\cdot\hat{\boldsymbol{\mu}}_{+}$, $\hat{v}^B_1=\openone$ and
$F=\mathbf{e}^{*}\cdot\hat{\boldsymbol{\mu}}_{-}$. 
The numerator of $\mathcal{I}_1$ simplifies to $|| \widetilde{V}_1\widetilde{\psi} (t_1,\mathbf{z}^{*})||^2=1+\mu_{\mathrm{eff}}^2$, with  $\mu_{\mathrm{eff}}^2= \sum_n(\mathbf{e}\cdot{\boldsymbol{\mu}}_n)^2$.
We calculate the final state $\ket{\widetilde{\psi}(t,z^{*})}$, defined in Eq.~(\ref{eq:def_double}), following the prescription in Section \ref{sec:numeric_prop}:
After the interaction of the initial vector $\ket{\widetilde{\psi}(t_1,\mathbf{z}^{*})}$ with the operator $\widetilde{V}_1$  the state becomes 
$\begin{pmatrix}
\ket{g}\\
(\mathbf{e}\cdot\hat{\boldsymbol{\mu}}_{+})|g\rangle
\end{pmatrix}
$.
During the subsequent time propagation (from $t_1$ to $t$), the \textit{bra} is in the ground state and  only acquires a phase, $\ket{\phi_\mathrm{B}(t)}=e^{-i \epsilon_\mathrm{g}t}\ket{g}$.
The \textit{ket} contribution $\ket{\phi_\mathrm{K}(t)}=\ket{\phi_\mathrm{K}(t,\mathbf{z}^*)}$ can be obtained from propagating the initial state $(\mathbf{e}\cdot\hat{\boldsymbol{\mu}}_{+})|g\rangle$ with the NMQDS equation in the {\it single} Hilbert space, where the expectation values of $\hat{L}_n^{\dagger}$ at time $s$ ($\langle \hat{L}_n^\dagger \rangle_s$) are calculated with respect to the norm $ ||\widetilde\psi(s,\mathbf{z}^{*})||^2=(||\phi_\mathrm{K}(s,\mathbf{z}^*)||^2 +1)$ of the state in the doubled Hilbert space 
\begin{equation}
    \langle \hat{L}_n^\dagger \rangle_s= \bra{\phi_\mathrm{K}(s,\mathbf{z}^*)}\hat{L}_n^\dagger\ket{\phi_\mathrm{K}(s,\mathbf{z}^*)}/(||\phi_\mathrm{K}(s,\mathbf{z}^*)||^2 +1).
\end{equation}

Finally the response function can be written as 
\begin{equation}
    R^{(1)}(t)= \mathcal{M} \Big[\frac{1+\mu_{\mathrm{eff}}^2}{||\phi_\mathrm{K}(t,\mathbf{z}^{*})||^2+1} \mu_{\mathrm{eff}}\langle\psi_{\mathrm{ex}}|\phi_\mathrm{K}(t,\mathbf{z}^{*})\rangle\Big]e^{i\epsilon_gt},
\end{equation}
where we have introduced $\ket{\psi_\mathrm{ex}}= \frac{1}{\mu_{\mathrm{eff}}} \mathbf{e}\cdot\hat{\boldsymbol{\mu}}_{+}\ket{g}$,
to make the connection to our previous result\cite{ODHOPSLP} more obvious (see next subsection). 

\subsection{Relation to previous results}
In a previous publication\cite{ODHOPSLP} we have derived an equation for the perturbative calculation of the linear response function using the non-linear  NMQSD equation.
In that work the starting point was to treat the non-Hermitean operator $(\mathbf{e}\cdot\hat{\boldsymbol{\mu}}_{+})|g\rangle\langle{g}|\hat{\rho}_{\mathrm{B}}$ as `initial state', which is then decomposed into a sum of pure states which can be propagated via NMQSD.
In  Ref.~\onlinecite{ODHOPSLP} the response was obtained from
\begin{equation}
\label{eq:R_decomp}
    R_\mathrm{decomp}(t)= \mu_\mathrm{eff}^2 \mathcal{M}\frac{\braket{\psi_\mathrm{ex}}{\chi(t,\mathbf{z}^*)}}{\frac{1}{2}(||\chi(t,\mathbf{z}^*)|| +1)}e^{i \epsilon_g t}.
\end{equation}
Also here the state $\ket{\chi}$ is propagated in the excited Hilbert space, but expectation values of $\hat{L}_n^{\dagger}$ are calculated using the normalization with $(||\chi(t,\mathbf{z}^*)|| +1)$.
From this one sees that the only difference to the approach of section \ref{sec:ResponseInNMQSD} is that one starts the excited state propagation of the \textit{ket} with a different normalized state.
The change in initial condition leads to different trajectories even for the same noise realization. Nevertheless, both methods result in equivalent average response functions.
Numerically we have found that for our examples there is little difference in the convergence of the two approaches.

To derive Eq.~(\ref{eq:R_decomp}) within our present formalism we redefine the response function (section \ref{sec:ResponseFunct}) by scaling 
$\hat{V}_j\rightarrow \hat{V}_j/ m_j$, $\hat{F}\rightarrow \hat{F}/m$ and $R^{(M)}\rightarrow m\, (\prod_{j}^M m_j)\, R^{(M)} $. Using $m=\mu_\mathrm{eff}$ and $m_j=\mu_\mathrm{eff}$ we have explicitly for the response function (\ref{eq:Resp_LinAbs})
\begin{equation}
    R^{(1)}(t)=\mu_{\mathrm{eff}}^2\mathrm{Tr}\left\lbrace\frac{(\mathbf{e}^{*}\cdot\hat{\boldsymbol{\mu}}_{-})}{\mu_{\mathrm{eff}}}e^{-i\hat{H}t}\frac{(\mathbf{e}\cdot\hat{\boldsymbol{\mu}}_{+})}{\mu_{\mathrm{eff}}}|g\rangle\langle{g}|\hat{\rho}_{\mathrm{B}}e^{i\hat{H}t}\right\rbrace.
\end{equation}
We then use the same steps as in the previous subsection and arrive at Eq.~(\ref{eq:R_decomp}).

\section{The error measure\label{sec:error}}
To quantify the difference between different 2D spectra we introduce an error measure in the following way:
First we normalize each spectrum according to
\begin{equation}
   S(\omega_t,\omega_\tau) \rightarrow S(\omega_t,\omega_\tau)/S
\end{equation}
with
\begin{equation}
    S=\frac{1}{\Omega} \int_{\omega_\mathrm{min}}^{\omega_\mathrm{max}} d\omega_t \int_{\omega_{\mathrm{min}}}^{\omega_\mathrm{max}} d \omega_{\tau}|S (\omega_t,\omega_\tau)|
\end{equation}
with $\Omega=1$.
For two spectra $S^{(1)}(\omega_t,\omega_\tau)$ and $S^{(2)}(\omega_t,\omega_\tau)$ we then introduce the difference
\begin{equation}
     \Delta S(\omega_t,\omega_\tau)= S^{(1)}(\omega_t,\omega_\tau)
     -S^{(2)}(\omega_t,\omega_\tau)
\end{equation}
Finaly, we define the {\it integrated difference}
\begin{equation}
   \mathbb{E} =
    \int_{\omega_\mathrm{min}}^{\omega_\mathrm{max}} d\omega_t \int_{\omega_\mathrm{min}}^{\omega_\mathrm{max}} d \omega_{\tau} |\Delta S (\omega_t,\omega_\tau)|
\end{equation}

To obtain a detailed analysis of the statistical error due to a finite number of trajectories shown in Figs.~\ref{fig:Weak} and \ref{fig:Strong}, we employ bootstrapping.\cite{BootStrap} We first calculate  $4\times{10}^4$ trajectories. For each value of $N_{\mathrm{traj}}$, we then construct 500 ensembles by randomly choosing $N_{\mathrm{traj}}$ trajectories from the original $4\times{10}^4$ trajectories. For each ensemble, we calculate the averaged {\it integrated difference} and finally obtain the {\it error} as the mean of the {\it integrated difference} over the 500 ensembles.


\end{document}